\documentclass[12pt]{iopart}
\usepackage{graphicx}
\usepackage{ulem}
\usepackage{color}
\usepackage{iopams}
\usepackage{setstack}
\usepackage{mathabx}

\begin{document}

\title{Bayesian mechanics of perceptual inference and motor control in the brain}

\author{Chang Sub Kim}
\address{Department of Physics,
Chonnam National University,
Gwangju 61186, Republic of Korea}
\ead{cskim@jnu.ac.kr}

\begin{abstract}
The free energy principle (FEP) in the neurosciences stipulates that all viable agents induce and minimize informational free energy in the brain to fit their environmental niche.
In this study, we continue our effort to make the FEP a more physically principled formalism by implementing free energy minimization based on the principle of least action.
We build a Bayesian mechanics (BM) by casting the formulation reported in the earlier publication (Kim in Neural Comput 30:2616–2659, 2018) to considering active inference beyond passive perception.
The BM is a neural implementation of variational Bayes under the FEP in continuous time.
The resulting BM is provided as an effective Hamilton's equation of motion and subject to the control signal arising from the brain's prediction errors at the proprioceptive level.
To demonstrate the utility of our approach, we adopt a simple agent-based model and present a concrete numerical illustration of the brain performing recognition dynamics by integrating BM in neural phase space.
Furthermore, we recapitulate the major theoretical architectures in the FEP by comparing our approach with the common state-space formulations.

\end{abstract}

\vskip1in
\maketitle

\section{Introduction}
The free energy principle (FEP) in the field of neurosciences rationalizes that all viable organisms cognize and behave in the natural world by calling forth the probabilistic models in their neural system {\textemdash} the brain {\textemdash} in a manner that ensures their adaptive fitness (Friston 2010a).
The neurobiological mechanism that endows an organism's brain {\textemdash} the neural observer {\textemdash} with this ability is theoretically framed into an inequality that weighs two information-theoretical measures: surprisal and informational free energy (IFE) (see, for a review, Buckley and Kim et al. 2017).
The surprisal provides a measure of the atypicality of an environmental niche, and the IFE is the upper bound of the surprisal.
The inequality enables a cognitive agent to minimize the IFE as a variational objective function indirectly instead of the intractable surprisal.\footnote{Free energy (FE) is a notion developed by Hermann von Helmholtz in thermodynamics; it is a physical energy measured in joules. The FE in the FEP is an information-theoretic measure defined in terms of probabilities, which serves as an objective function for variational Bayesian inference. Accordingly, we call it variational IFE in our formulation.}
The minimization corresponds to inferring the external causes of afferent sensory data, which are encoded as a probability density at the sensory interface, e.g., sensory organs.
The brain of an organism neurophysically performs the Bayesian computation of minimizing the induced variational IFE; this is termed as recognition dynamics (RD), which emulates, under the Laplace approximation (Friston et al. 2007), the predictive coding scheme of message processing or recognition (Rao and Ballard 1999; Bogacz2017).
The neuronal self-organization \textit{in vitro} under the FEP was studied recently at the level of single neuron responses (Isomura et al. 2015; Isomura and Friston 2018).
Owing to its explanatory power of perception, learning, and behavior of living organisms within a framework, it is suggested a promising unified biological principle (Friston 2010a; Friston 2013; Colombo and Wright 2018; Ramstead et al. 2018).

The neurophysical mechanisms of the abductive inference in the brain are yet to be understood; therefore, researchers mostly rely on information-theoretic concepts (Elfwing 2016; Ramstead et al. 2019; Kuzma 2019; Shimazaki 2019; Kiefer 2020, Sanders et al. 2020).
The FEP facilitates dynamic causal models in the brain's generalized-state space (Friston 2008b; Friston et al. 2010b), which pose a mixed discrete-continuous Bayesian filtering (Jazwinski 1970; Balaji and Friston 2011).
In this work, we consider that the brain confronts the continuous influx of stochastic sensations and conducts the Bayesian inversion of inferring external causes in the continuous state representations.
Biological phenomena are naturally continuous spatiotemporal events; accordingly, we suggest that the continuous-state approaches used to describe cognition and behavior are better suited than discrete-state descriptions for studying the perceptual computation in the brain.

Recently, we carefully evaluated the FEP while clarifying technical assumptions that underlie the continuous state-space formulation of the FEP (Buckley and Kim et al. 2017).
A full account of the discrete-state formulation complementary to our formulation can be found in (Da Costa et al. 2020a).
In a subsequent paper (Kim 2018), we reported a different variational scheme that the Bayesian brain may utilize in conducting inference.
In particular, by postulating that ``surprisal'' plays the role of a Lagrangian in theoretical mechanics (Landau and Lifshitz 1976; Sengupta et al. 2016), we worked a plausible computational implementation of the FEP by utilizing the principle of least action.
We believed that although the FEP relies on Bayesian abductive computation, it must be properly formulated conforming to the physical principles and laws governing the matter comprising the brain.
To this end, we proposed that any process theory of the FEP ought to be based on the full implication of the inequality (Kim 2018)
\begin{equation}\label{FEP}
\int dt\left\{-\ln p(\varphi)\right\} \le \int dt{\cal F}[q(\vartheta),p(\varphi,\vartheta)],
\end{equation}
where $\varphi$ and $\vartheta$ collectively denote the sensory inputs and environmental hidden states, respectively.
The integrand on the left-hand side (LHS) of the preceding equation $-\ln p(\varphi)$ is the aforementioned \textit{surprisal}, which measures the ``self-information'' contained in the sensory density $p(\varphi)$ (Cover 2006),
and $\cal F$ on the right-hand side (RHS) is the variational IFE defined as
\begin{equation}
{\cal F}[q(\vartheta),p(\varphi,\vartheta)] \equiv \int d\vartheta q(\vartheta)\ln\frac{q(\vartheta)}{p(\varphi,\vartheta)}, \label{IFE}
\end{equation}
which encapsulates the recognition (R-) density $q(\vartheta)$ and the generative (G-) density $p(\vartheta,\varphi)$ (Buckley and Kim et al. 2017).
While the G-density represents the brain's belief (or assumption) of sensory generation and hidden environmental dynamics,
the R-density is the brain's current estimate of the environmental cause of the sensory perturbation.
The G- and R- densities together induce variational IFE when receptors at the brain-environment interface are excited by sensory perturbations.

According to Eq.~(\ref{FEP}), the FEP articulates that the brain minimizes the upper bound of the sensory uncertainty, which is a long-term surprisal.
We identify this bound as an informational action (IA) within the scope of the mechanical principle of least action (Landau and Lifshitz 1976).
Then, by complying with the revised FEP, we formulate the Bayesian mechanics (BM) that executes the RD in the brain.
The RD neurophysically performs the computation for minimizing the IA when the neural observer encounters continuous streams of sensory data.
The advantage of our formulation is that the brain and the environmental states are specified using only bare continuous variables and their first-order derivatives (velocities or equivalent momenta).
The momentum variables represent prediction errors, which quantify the discrepancy between an observed input and its top-down belief of a cognitive agent in the predictive coding language (Huang and Rao 2011; de Gardelle et al. 2013; Kozunov et al. 2020).

The goal of this work is to cast our previous study to include the agent's motor control, which acts on the environment to alter sensory inputs.\footnote{In this work, we use the term \textit{control} instead of the frequently used term ``action'' to mean the motion of a living agent's effectors (muscles) acting on the environment.
This is done to avoid any confusion with the term \textit{action} appearing in the nomenclature of ``the principle of least action''.}
Previously, by utilizing the principle of least action, we focused on the formulation of perceptual dynamics for the \textit{passive inference} of static sensory inputs (Kim 2018) without incorporating motor control for the active perception of nonstationary sensory streams.
Here, we apply our approach to the problem of \textit{active inference} derived from the FEP (Friston et al. 2009; Friston et al. 2010c; Friston et al. 2011a), which proposes that organisms can minimize the IFE by altering sensory observations when the outcome of perceptual inference alone is not in accordance with the internal representation of the environment (Buckley and Kim et al. 2017).
Living systems are endowed with the ability to adjust their sensations via proprioceptive feedback, which is attributed to an inherited trait of all motile animals embodied in the reflex pathways (Tuthill and Azim 2018).
In this respect, motor control is considered an inference of the causes of motor signals encoded as prediction errors at proprioceptors, and motor inference is realized at the spinal level by classical reflex arcs (Friston 2011b; Adams et al. 2013).
Our formulation evinces time-dependent driving terms in the obtained BM, which arise from sensory prediction errors, as control (motor) signals.
Accordingly, the BM bears resemblance to the deterministic control derived from Pontryagin’s maximum principle in optimal control theory (Todorov 2007).
In this work, we consider the agent's locomotion for action inference only implicitly:
our formulation focuses on the implementation of the control signal (or commands) at the neural level of description and not at the behavioral level of biological locomotion;
accordingly, the additional minimization mechanism of the IA inferring optimal control was not explicitly handled, which is left as a future work.
There are other systematic approaches that try to relate the active inference formalism to the existing control theories (Baltieri and Buckley 2019; Millidge et al. 2020a; Da Costa et al. 2020c).

Technically, a variation in the IA yields the BM that computes Bayesian inversion, which is given as a set of coupled differential equations for the brain variables and their conjugate momenta.
The brain variables are ascribed to the brain's representation of the environmental states, and their conjugate momenta are the combined prediction errors of the sensory data and the rate of the state representations.
The neural computation of active inference corresponds to the BM integration and is subject to nonautonomous motor signals.
The obtained solution results in optimal trajectories in the perceptual phase space, which yields a minimum accumulation of the IFE over continuous time, i.e., a stationary value of the IA.
Our IA is identical to that of “free action” defined in the Bayesian filtering schemes (Friston et al. 2008c).
When the minimization of free action is formulated in the generalized filtering scheme (Friston et al. 2010b), two approaches are akin to each other such that both assume the Laplace-encoded IFE as a mechanical Lagrangian.
The difference lies in their mathematical realization of minimization: our approach applies the principle of least action in classical mechanics, while generalized filtering uses the gradient descent method in the generalized state space, where generalized states are interpreted as a solenoidal gradient flow in a nonequilibrium steady state.

The remainder of this paper is organized as follows.
In Sect.~\ref{synopsis}, we unravel some of the theoretical details in the formulation of the FEP.
In Sect.~\ref{active-inference}, we formulate the BM of the sensorimotor cycle by utilizing the principle of least action.
Then, in Sect.~\ref{numerics}, we present a parsimonious model for a concrete manifestation of our formulation.
Finally, in Sect.~\ref{conclusion}, we provide the concluding remarks.

\section{Recapitulation of technical developments}
\label{synopsis}
Here, we recapitulate theoretical architectures in the continuous-state formulation under the FEP while discussing technical features that distinguish our formulation from prevailing state-space approaches.

\subsection{Perspective on generalized states}
\label{generalized states}
The Bayesian filtering formalism of the FEP adopts the concept of the generalized motion of a dynamical object by defining its mechanical state beyond position and velocity (momentum).
The generalized states of motion are generated by recursively taking time derivatives of the bare states.
A point in the hyperspace defined by the generalized states is interpreted as an instantaneous trajectory.
This notion provides an essential theoretical basis for ensuring an equilibrium solution of the RD in the conventional formulation of the FEP (Kim 2018); it is commonly employed by researchers (Parr and Friston 2018; Baltieri and Buckley 2019).

The motivation behind the generalized coordinates of motion is to describe noise correlation in the generative processes beyond white noise (Wiener process), and thus, to provide a more detailed specification of the dynamical states (Friston 2008a; Friston 2008b; Friston et al. 2010b).
The mathematical theory of a quasi-Markovian process undergirds this formulation, which describes general stochastic dynamics with a colored-noise correlation with a finite-dimensional Markovian equation in an extended state space by adding auxiliary variables (Pavliotis 2014).
State-space augmentation in terms of generalized coordinates may be considered a special realization of the Pavliotis formalism.
The state-extension procedure adopts some specific approximations, such as the local linearization procedure developed in nonlinear time-series analysis (Ozaki 1992).

From the physics perspective, higher-order states possess a different dynamical status in comparison with Newtonian mechanical states, specified only by position (bare order) and velocity (first order).
A change in the Newtonian states is caused by a force that specifies acceleration (second order) (Landau and Lifshitz 1976).
Although there are no ``generalized forces'' causing the jerk (third order), snap (fourth order) etc., the jerk can be measured phenomenologically by observing a change in acceleration.
This induces all higher-order states to the kinematic level.
Another perspective is whether update equations in terms of generalized coordinates are equivalent to the Pavliotis' quasi-Markovian description.
Auxiliary variables in Pavliotis’ analysis are not generated by the recursive temporal derivatives of a bare state.
The generalized phase space considered in (Kerr and Graham 2000) is also spanned in terms of canonical displacement and momentum variables.
A further in-depth analysis is required.

Our formulation does not employ the generalized states, but instead, it follows the normative rules in specifying generative models (Kim 2018).
The derived BM performs the brain’s Bayesian inference in terms of only the bare brain variable and its conjugate momentum in phase space, and not in an extended state space.
Accordingly, our formulation is restricted to the white noise in the generative processes; however, it provides a natural approach to determine the equilibrium solutions of the BM (see Sect.~\ref{numerics}).
For the general brain models described by many brain variables, the brain's BM can be set up in multi-dimensional phase space, which is distinctive from the state-space augmentation in the generalized coordinate formulation (see Sect.~\ref{niose correlation}).

\subsection{Continuous state implementation of recognition dynamics (RD)}
\label{replacing gradient descent}
The conventional FEP employs the gradient-descent minimization of the variational IFE by the brain's internal states.
To incorporate the time-varying feature of sensory inputs, the method distinguishes the path of a mode and the mode of a path in the generalized state space (Friston 2008b; Friston et al. 2008c; Friston et al. 2010b).
This theoretical construct intuitively considers the nonequilibrium dynamics of generalized brain states as drift-diffusion flows that locally conserve the ensemble density in the hyperspace of the generalized states (Friston and Ao 2012b; Friston 2019).

Mathematically, the gradient descent formulation is based on the general idea for a fast and efficient convergence and it ensures that formulations reach a sophisticate level by incorporating the Riemannian metric in information geometry (Amari 1998; Surace et al. 2020); the idea is applied to the FEP (Sengupta and Friston 2017, Da Costa et al. 2020b).

In our proposed formulation, we replace the gradient descent scheme with the standard mechanical formulation of the least action principle (Kim 2018).
However, there is a disadvantage in that we incorporate only the Gaussian white noise in the generative processes of the sensory data and environmental dynamics [see Sect.~\ref{niose correlation}].
The resulting novel RD described by an effective Hamiltonian mechanics entails optimal trajectories but no single fixed points in the canonical state (phase) space, which provides an estimate of the minimum sensory uncertainty, i.e., the average surprisal over a finite temporal horizon.
The phase space comprises the positions (predictions) and momenta (prediction errors) of the brain's representations of the causal environment.

Our implementation of the minimization procedure is an alternative to the gradient descent algorithms in the FEP.
A crucial difference between the two approaches is that while the gradient descent scheme searches for an instantaneous trajectory representing a local minimum on the IFE landscape in the multidimensional generalized state space, our theory determines an optimal trajectory minimizing the continuous-time integral of the IFE in two-dimensional phase space for a single variable problem.

\subsection{Treatment of noise correlations}
\label{niose correlation}
The FEP requires the brain's internal model of the G-density $p(\varphi,\vartheta)$ encapsulating the likelihood $p(\varphi|\vartheta)$ and prior $p(\vartheta)$.
The likelihood density is determined by the random fluctuation in the expected sensory-data generation, and the prior density is determined by that in the believed environmental dynamics.
The brain encounters sensory signals on a timescale, which is often shorter than the correlation time of the random processes (Friston 2008a); accordingly, in general, the noises embrace a non-Markovian stochastic process with an intrinsic temporal correlation that surmounts the ideal white-noise stochasticity.
Conventional formulations (Friston 2008b; Friston et al. 2010b) consider that colored noises are analytic (i.e., differentiable) to allow correlation between the distinct dynamical orders of the continuous states.
In practice, to furnish a closed dynamics for a finite number of variables, the recursive equations of motion for the continued generalized states need to be truncated at an arbitrary embedding order.

Our formulation considers the BM in the brain in terms of the standard Newtonian (Hamiltonian) construct; the drawback is that our theory does not explore the nature of temporal correlation in the assumed Gaussian noises in the generative processes.
Accordingly, our generative models assume and account for the white noise describing the Wiener processes.
The delta-correlated white noise is mathematically singular; they need to be smoothed to describe fast biophysical processes.
There are approaches in stochastic theories that formulate non-Markovian processes with colored noises without resorting to generalized states of motion (van Kampen:1981; Fox 1987; Risken 1989; Moon and Wettlaufer 2014), which are not discussed here.

Instead, we discuss an approach to extend the phase-space dimension for the white noise processes.
At the level of the Hodgkin\textendash Huxley description of the biophysical brain dynamics, the membrane potential, gating variables, and ionic concentrations are relevant coarse-grained brain variables (Hille 2001).
Thus, if one employs the fluctuating Hodgkin-Huxley models with Gaussian white noises as neurophysically plausible generative models (Kim 2018), one can proceed with our Lagrangian (equivalently, Hamiltonian) approach to formulate the RD in an extended phase space.
Such a state-space augmentation is different from and alternative to that in terms of the generalized coordinates of motion [see Sect.~\ref{generalized states}], while accommodating only delta-correlated noises.

\subsection{Lagrangian formulation of Bayesian mechanics (BM)}
\label{Bayesian mechanics}
The (classical) ``action'' is defined as an \textit{ordinary time-integral} of the Lagrangian for an arbitrary trajectory (Landau and Lifshitz 1976).
Our formulation of the BM proposes the Laplace-encoded IFE {\textemdash} an upper bound on the sensory surprisal {\textemdash} as an informational Lagrangian and hypothesizes the time-integral of the IFE {\textemdash} an upper bound on the sensory Shannon uncertainty {\textemdash} as an informational action (IA).
By applying the principle of least action, we minimize the IA to find a tight bound for the sensory uncertainty and derive the BM that performs the brain's Bayesian inference of the external cause of sensory data.
In turn, we cast the working BM in our formulation as effective Hamilton’s equations of motion in terms of position and momentum in phase space.

Meanwhile, the BM described in (Friston 2019) intuitively adopts the idea of Feynman’s path integral formulation (Feynman and Hibbs 2005).
The Feynman’s path integral formulation extends the idea of classical action to quantum dynamics and provides an approach to determine the ``propagator'' that specifies the transition probability between initial and final states.
The propagator is defined as a \textit{functional integral} of the exponentiated action, which summates all possible trajectories connecting initial and final states.
The description provided in (Friston 2019) identifies the propagator using the probability density over neural states, and it makes the connection to the Bayesian FEP.
In this manner, the surprisal may be identified as a negative log of the steady-state density in nonequilibrium ensemble dynamics (Parr et al. 2020), which is governed by a Fokker-Plank equation.
The generalized Bayesian filtering scheme (Friston et al. 2008c; Friston et al. 2010b) provides a continuous-state formulation of minimizing the surprisal, and it delivers the BM in terms of the generalized coordinates of motion using the concept of gradient flow.

In some technical details, the Lagrangian presented in (Friston 2019), which is the integrand in the classical action, encloses two terms.
They are the quadratic term arising from the state equation and the term involving a state-derivative (divergence in three dimension) of the force, which appears in the Langevin-type state equation.
The former term is included in our Lagrangian but with an additional quadratic term from the observation equation.
In contrast, the latter is not present in our Lagrangian, which is known to arise from the Stratonovich convention (Seifert 2012; Cugliandolo and Lecomte 2017).

\subsection{Closure of the sensorimotor loop in active inference}
The conventional FEP facilitates gradient descent minimization for the mechanistic implementation of active inference, which makes the motor-control dynamics available in the brain's RD (Friston et al. 2009; Friston et al. 2010c; Friston et al. 2011a).
The gradient-descent scheme is mathematically expressed as
\begin{equation}
\label{action}
\dot a = - \nabla_a {F} \rightarrow -\frac{\partial{F}}{\partial\varphi} \frac{d\varphi}{da},
\end{equation}
where $a$ denotes an agent's motor variable, and $F$ represents the Laplace-encoded IFE by the biophysical brain variables (Buckley and Kim et al. 2017).
An agent's capability of subjecting sensory inputs to motor control is considered a functional dependence $\varphi=\varphi(a)$ in the environmental generative processes (Friston et al. 2009).
According to Eq.~(\ref{action}), an agent performs the minimization by effectuating the sensory data $d\varphi/da$ and obtains the best result for motor inference when $\dot a=0$, where the condition $\partial{F}/\partial\varphi=0$ must be met.
Because $\partial{F}/\partial\varphi$ produces terms proportional to the sensory prediction errors, the fulfillment of motor inference is equivalent to suppressing proprioceptive errors.
Thus, motor control attempts to minimize prediction errors, while prediction errors convey motor signals for control dynamics; this forms a sensorimotor loop.
Some subtle questions arise here regarding the dynamical status of the motor-control variable $a$:
Equation~(\ref{action}) evidently handles $a$ as a dynamical state; however, the corresponding equation of motion governing its dynamics is not given in the environmental processes.
Instead, the mechanism of motor control that vicariously alters the sensory-data generation is presumed (Friston et al. 2009).
In addition, motor variables are represented as the active states of the brain, e.g., motor-neuron activities in the ventral horn of the spinal cord (Friston et al. 2010c); however, they are treated differently from other hidden-state representations.
Recall that the internal state representations are expressed as generalized states, whereas the active states are not.

In the following, we pose a semi-active inference problem that does not explicitly address optimal motor control (motor inference) in the RD but encompasses the motor-control signal as a time-dependent driving term arising from nonstationary prediction errors in the sensory-data cause.

\section{Closed-loop dynamics of perception and motor control}
\label{active-inference}
The brain is not divided into sensory and motor systems.
Instead, it is one inference machine that performs the closed-loop dynamics of perception and motor control.
Here, we develop a framework of active inference within the scope of the least action principle by employing the Laplace-encoded IFE as an informational Lagrangian.

The environmental states $\vartheta$ undergo deterministic or stochastic dynamics by obeying physical laws and principles.
Here, we do not explicitly consider their equations of motion because they are hidden from the brain's perspective, i.e., the brain as a neural observer does not possess direct epistemic access.
Similarly, sensory data $\varphi$ are physically generated by an externally hidden process at a sensory receptor, which constitutes the brain-environment interface.
However, to emphasize the effect of an agent's motor control $a$ on sensory generation, we facilitate the generative process of sensory data using an instantaneous mapping
\begin{equation}
\label{ex-obseq}
\varphi=h(\vartheta,a)+z_{gp},
\end{equation}
where $h(\vartheta,a)$ denotes the linear or nonlinear map of input generation, and $z_{gp}$ represents the noise involved.
Note that an agent's motor-control $a$ is explicitly included in the generative map.
However, the neural observer is not aware of how the sensory streams are effectuated by the agent's motion in the environment (Friston et al. 2010c).

The FEP circumvents this epistemic difficulty by hypothesizing a formal homology between external physical processes and the corresponding internal models foreseen by the neural observer (Friston et al. 2010c).
Upon receiving sensory-data influx, the brain launches R-density $q(\vartheta)$ to infer the external causes via variational Bayes.
The R-density is the probabilistic representation of the environment, whose sufficient statistics are assumed to be encoded by neurophysical brain variables, e.g., neuronal activity or synaptic efficacy.
When a fixed-form Gaussian density is considered for the R-density, which is called Laplace approximation, only the first-order sufficient statistic, i.e., the mean $\mu$ is needed to specify the IFE effectively (Buckley and Kim et al. 2017).
The brain continually updates the R-density using its internal dynamics, described here as a Langevin-type equation
\begin{equation}
\label{state-eq}
\frac{d\mu}{dt}=f(\mu)+w,
\end{equation}
where $f(\mu)$ represents the brain's belief regarding the external dynamics encoded by a neurophysical driving mechanism of the brain variables $\mu$, and $w$ is random noise.
The sensory perturbations at the receptors are predicted by the neural observer via the instantaneous mapping
\begin{equation}
\label{obs-eq}
\varphi(a)=g(\mu)+z,
\end{equation}
where the belief $g(\mu)$ is encoded by the internal variables, and $z$ is the associated noise.
Our sensory generative model provides a mechanism for sampling sensory data $\varphi$ using the brain's active states $a$, which represent an external motor control embedded in Eq.~(\ref{ex-obseq}).
Note that Eq.~(\ref{ex-obseq}) describes the environmental processes that generate sensory inputs $\varphi$, while its homolog ``Eq.~(\ref{obs-eq})'' prescribes the brains' prior belief of $\varphi$ that can be altered by the active states $a$.
The instantaneous state of the brain $\mu$, which is specified by Eq.~(\ref{state-eq}), selects a particular R-density $q(\vartheta)$ when the brain seeks the true posterior (the goal of perceptual inference).
The motor control fulfills the prior expectations by modifying the sensory generation via active-state effectuation at the proprioceptors.

Through Laplace approximation (Buckley and Kim et al. 2017), the G-density $p(\varphi,\vartheta)$ is encoded in the brain as $p=p(\varphi,\mu)$, where the sensory stimuli $\varphi$ are predicted by the neural observer $\mu$ via Eq.~(\ref{obs-eq}).
Here, we argue that the physical sensory-recording process is conditionally independent of the brain's internal dynamics; however, the brain states must be neurophysically involved in computing the sensory prediction.
In other words, from the physics perspective, the sensory perturbation $\varphi$ at the interface is a source for exciting the neuronal activity $\mu$.
This observation renders the set of Eqs.~(\ref{state-eq}) and (\ref{obs-eq}) to be dynamically coupled, and not conditionally independent.
We incorporate this conditional dependence into our formulation by introducing a statistical coupling via the covariance connection between the likelihood $p(\varphi|\mu)$ and prior $p(\mu)$ that together furnish the Laplace-encoded G-density.

For simplicity, we consider the stationary Gaussian processes for the bivariate variable $Z$ as a column vector
\[
Z \equiv
\left(\begin{array}{c}
w  \\
z
\end{array}\right),
\]
where $w=\dot\mu-f(\mu)$ and $z=\varphi-g(\mu)$, and we specify the Laplace-encoded G-density $p(\varphi,\mu)=p(\varphi|\mu)p(\mu)$ as
\begin{equation}
\label{expecedG}
p(\varphi,\mu) = \frac{1}{\sqrt{(2\pi)^2|\Sigma|}}\exp\left(-\frac{1}{2}Z^T\Sigma^{-1}Z\right),
\end{equation}
where $|\Sigma|$ and $\Sigma^{-1}$ are the determinant and the inverse of the matrix $\Sigma$, respectively; $Z^T$ is the transpose of $Z$.
The covariance matrix $\Sigma$ for the above is given as
\[
\Sigma=
\left(\begin{array}{cc}
\sigma_w & \phi(t)\\
\phi(t) & \sigma_z
\end{array}\right),
\]
where the stationary variances $\sigma_i$ ($i=w,z$) and the transient covariance $\phi$ are defined, respectively, as
\[\sigma_w(0)=\langle w^2\rangle,\ \sigma_z(0)=\langle z^2\rangle,\ {\rm and}\ \phi(t)=\langle w(0)z(t)\rangle.
\]
With the prescribed internal model of the brain for the G-density, the Laplace-encoded IFE can be specified as $F(\varphi,\mu)=-\ln p(\varphi,\mu)$ (for details, see Buckley and Kim et al. 2017).
Then, it follows that
\begin{eqnarray}
\label{Laplace-IFE}
F(\varphi,\mu;t) &=& \frac{1}{2}m_w\left(\dot\mu-f(\mu)\right)^2 +\frac{1}{2}m_z\left(\varphi-g(\mu)\right)^2 \\ &&-\sqrt{m_wm_z}\rho\left(\dot\mu-f(\mu)\right)\left(\varphi-g(\mu)\right)\nonumber\\
&&+\frac{1}{2}\ln\left( 2\pi(1-\rho^2)\sigma_w\sigma_z\right),\nonumber
\end{eqnarray}
where $\rho$ denotes the correlation function defined as a normalized covariance
\begin{equation}\label{correlation}
\rho\equiv\frac{\phi}{\sqrt{\sigma_w\sigma_z}}.
\end{equation}
Furthermore, we introduce notations $m_i\ (i=w,z)$ as
\begin{equation}\label{neural-mass}
m_i\equiv\frac{1}{\sigma_i(1-\rho^2)},
\end{equation}
which are \textit{precisions}, scaled by the correlation in the conventional FEP.

Next, as proposed in (Kim 2018), we identify $F$ as an informational Lagrangian $L$ within the scope of the principle of least action, and we define
\begin{eqnarray}
\label{Lagrangian}
L &\equiv& \frac{1}{2}m_w\left(\dot\mu-f(\mu)\right)^2 +\frac{1}{2}m_z\left(\varphi-g(\mu)\right)^2 \nonumber\\ &&-\sqrt{m_wm_z}\rho\left(\dot\mu-f(\mu)\right)\left(\varphi-g(\mu)\right),
\end{eqnarray}
which is viewed as a function of $\mu$ and $\dot\mu$ for the given sensory inputs $\varphi(t)$, i.e., $L=L(\mu,\dot\mu;\varphi)$.
Note that we dropped the last term in Eq.~(\ref{Laplace-IFE}) when translating $F$ into $L$ because it can be expressed as a total time-derivative term that does not affect the resulting equations of motion (Landau and Lifshitz 1976).
Then, the theoretical action $S$ that effectuates the variational objective functional under the revised FEP is set up as
\begin{equation}
\label{IA}
S[\mu(t)]=\int F(\mu(t),\dot\mu(t);\varphi) dt.
\end{equation}
The Euler-Lagrange equation of motion, which determines the trajectory $\mu=\mu(t)$ for a given initial condition $\mu(0)$, is derived by minimizing the action $\delta S\equiv0$.

Equivalently, the equations of motion can be considered in terms of the position $\mu$ and its conjugate momentum $p$, instead of the position $\mu$ and velocity $\dot\mu$.
We used the terms position and velocity as a metaphor to indicate dynamical variables $\mu$ and $\dot\mu$, respectively.
For this, we need to convert Lagrangian $L$ into Hamiltonian $H$ by performing a Legendre transformation
\[H(\mu,p)=p\dot\mu-L(\mu,\dot\mu),\]
where $p$ denotes the canonical momentum conjugate to $\mu$, which is calculated from $L$ as
\begin{equation}
\label{momentum}
p=\frac{\partial L}{\partial \dot\mu}=m_w\left(\dot \mu-f\right)-\sqrt{m_wm_z}(\varphi-g)\rho.
\end{equation}
After some manipulation, the functional form of $H$ can be obtained explicitly as
\begin{equation}
\label{Ham1}
H(\mu,p;\varphi) = T(\mu,p;\varphi) + V(\mu;\varphi),
\end{equation}
where we indicated its dependence on the sensory influx $\varphi$.
In addition, the terms $T$ and $V$ on the RHS are defined as
\begin{equation}\label{kinetic}
T\equiv\frac{p^2}{2m_w}+\left(\rho\sqrt{\frac{m_z}{m_w}}\Big(\varphi-g(\mu)\Big)+f(\mu)\right)p,
\end{equation}
\begin{equation}\label{potential}
V\equiv\frac{1}{2}m_z\left(\rho^2-1\right)\Big(\varphi-g(\mu)\Big)^2.
\end{equation}
Here, $T$ and $V$ represent the \textit{kinetic and potential energies}, respectively, which define the informational Hamiltonian of the brain.
Similarly, $m_w$ and $m_z$ represent the \textit{neural inertial masse} as a metaphor.
Unlike that in standard mechanics, the second term in the expression for kinetic energy is dependent on linear momentum and position.

We generate the Hamilton equations of motion, which are equivalent to the Lagrange equation using
\[\dot\mu = \frac{{\partial H}}{ \partial p} \quad{\rm and}\quad \dot p = - \frac{{\partial H}}{ \partial \mu}.\]
As described below, Hamilton's equations are better suited for our purposes because they specify the RD as coupled first-order differential equations of the brain state $\mu$ and its conjugate momentum $p$.
In contrast, the Lagrange equation is a second-order differential equation of the state variable (Landau and Lifshitz 1976).
The results are
\begin{eqnarray}
\dot\mu &=& \frac{1}{m_w}p+f(\mu)+\alpha\Delta_\varphi, \label{HamEq1}\\
\dot p &=& - \left(\frac{\partial f}{\partial \mu}-\beta\frac{\partial g}{\partial\mu}\right)p -\left(1-\gamma^2\right)\frac{\partial g}{\partial\mu}\Delta_\varphi, \label{HamEq2}
\end{eqnarray}
where parameters $\alpha$, $\beta$, and $\gamma$ have been respectively defined for notational convenience as
\begin{equation}\label{parameters}
\alpha\equiv\frac{\rho}{\sqrt{m_zm_w}},\ \beta\equiv m_z\alpha,\ {\rm and}\ \gamma\equiv\rho\sqrt{\kappa},
\end{equation}
where $\kappa$ denotes the tuning parameter to spawn stability.
In Eqs.~(\ref{HamEq1}) and (\ref{HamEq2}), we defined the notation $\Delta_\varphi$ as
\begin{equation}
\label{MotorSignal}
\Delta_\varphi(\mu;t) \equiv m_z(\varphi(a)-g(\mu)).
\end{equation}
It measures the discrepancy between the adjustable sensory input $\varphi$ by an agent's motor control $a$ and the top-down neural prediction $g(\mu)$, weighted by the neural inertial mass $m_z$.

Below, we appraise the BM prescribed by Eqs.~(\ref{HamEq1}) and (\ref{HamEq2}) and note some significant aspects:
\begin{itemize}
\item[(i)] The derived RD suggests that both brain activities $\mu$ and their conjugate momenta $p$ are dynamic variables. The instantaneous values of $\mu$ and $p$ correspond to a point in the brain's perceptual phase space, and the continuous solution over a temporal horizon forms an optimal trajectory that minimizes the theoretical action, which represents sensory uncertainty.
\item[(ii)] The canonical momentum $p$ defined in Eq.~(\ref{momentum}) can be rewritten as $p=m_w\left(\dot \mu-f\right)-\rho\sqrt{m_w/m_z}\Delta_\varphi$. Accordingly, when the normalized correlation $\rho$ is nonvanishing, the momentum quantifies combined errors in predicting changing states and sensory stimuli.
    Prediction errors propagate through the brain by obeying coupled dynamics according to Eqs.~(\ref{HamEq1}) and (\ref{HamEq2}).
\item[(iii)] Terms involving time-dependent $\Delta_\varphi$ in Eqs.~(\ref{HamEq1}) and (\ref{HamEq2}) are identified as driving forces ${\cal C}_i$, $i=\mu,\ p$,
\begin{eqnarray}
&&{\cal C}_\mu\equiv \alpha\Delta_\varphi, \label{contriol1}\\
&&{\cal C}_p\equiv -\left(1-\gamma^2\right)\frac{\partial g}{\partial\mu}\Delta_\varphi. \label{control2}
\end{eqnarray}
The sensory prediction error $\Delta_\varphi$ defined in Eq.~(\ref{MotorSignal}) quantifies motor signals engaging the brain's nervous control in integrating the RD.
\end{itemize}

Equations~(\ref{HamEq1}) and (\ref{HamEq2}) are the highlights of our formulation, which prescribe the brain's BM of semi-actively inferring the external causes of sensory inputs under the revised FEP.
Note that the motor variable $a$ is not explicitly included in our derived RD; instead, it implicitly induces nonautonomous sensory inputs $\varphi(t)$ in the motor signal $\Delta_\varphi$.
The motor signal appears as a time-dependent driving force; accordingly, our Hamiltonian formulation bears a resemblance to the motor-control dynamics described by the Hamilton-Jacobi-Bellman (HJB) equation in the control theory (Todorov 2007).
If one regards the Lagrangian Eq.~(\ref{Lagrangian}) as a negative cost rate and the canonical momentum $p$ as a costate, our IA is equivalent to the total cost function that generates the continuous-state HJB equations.
In optimal control theory, the associated Hamiltonian function is further minimized with respect to the control signal, which we do not explicitly consider in this work.
In our formulation, the motor signals are produced by the discrepancy between the sensory streams $\varphi(t)$ and those predicted by the brain.
The nonstationary data are presented to a sensorimotor receptor, whose field position in the environment is specified by the agent's locomotive motion.
The neural observer continuously integrates the BM subject to a motor signal to perform the sensory-uncertainty minimization, thereby closing the perception and motion control within a reflex arc.
When we neglect the correlation $\rho$ between the sensory prediction modeled by Eq.~(\ref{obs-eq}) and the internal dynamics of predicting the neuronal state modeled by Eq.~(\ref{state-eq}), we can recover the RD reported in the previous publication (Kim 2018), which demonstrates the consistency of our formulation.

In the present treatment, we consider only a single brain variable $\mu$; accordingly, the ensuing BM specified by Eqs.~(\ref{HamEq1}) and (\ref{HamEq2}) is described in a two-dimensional phase space.
The extension of our formulation to the general case of the multivariate brain is possible by applying the same line of work proposed in (Kim 2018).
Under the independent-particle approximation, the multivariate Lagrangian takes the form
\begin{eqnarray}
\label{Lagrangian2}
L &\equiv& \frac{1}{2}\sum_{\alpha=1}^N{\Big[}m_{w\alpha}{\Big(}\dot\mu_\alpha-f_\alpha(\{\mu\}){\Big)}^2 +m_{z\alpha}{\Big(}\varphi_\alpha-g_\alpha(\{\mu\}){\Big)}^2 \nonumber\\
&-&2\rho_\alpha\sqrt{m_{w\alpha}m_{z\alpha}}\Big(\dot\mu_\alpha-f_\alpha(\{\mu\})\Big)
\Big(\varphi_\alpha-g_\alpha(\{\mu\})\Big){\Big]},
\end{eqnarray}
where $\{\mu\}=(\mu_1,\mu_2,\cdots,\mu_N)$ denotes a row vector of $N$ brain states that respond to multiple sensory inputs $\{\varphi\}=(\varphi_1,\varphi_2,\cdots,\varphi_N)$ in a general manner.
Note that our proposed multivariate formulation is different from the state-space augmentation using the higher-order states [see Sect.~\ref{generalized states}].
In our case, multiple brain states are, for instance, the membrane potential, gating variables, and ionic concentrations, that can be viewed as the fluctuating variables on a corase-grained time scale, influnced by Gaussian white noises [see Sect.~\ref{niose correlation}].

Furthermore, the implication of our formulation in the hierarchical brain can be achieved in a straightforward manner as in (Kim 2018), which adopts the bidirectional facet in information flow of descending predictions and ascending prediction errors (Markov and Kennedy 2013; Michalareas et al. 2016).
Note that in ensuing formulation, both descending predictions and ascending prediction errors will constitute the dynamical states governed by the closed-loop RD in the functional architecture of the brain's sensorimotor system.
This feature is in contrast to the conventional implementation of the FEP, which delivers the backward prediction {\textemdash} belief propagation {\textemdash} as neural dynamics and the forward prediction error as an instant message passing without causal dynamics (Friston 2010a; Buckley and Kim et al. 2017).

\section{Simple Bayesian-agent model: Implicit motor control}
\label{numerics}
In this section, we numerically demonstrate the utility of our formulation using an agent-based model, which is based on a previous publication (Buckley and Kim et al. 2017).
Unlike that in the previous study, the current model does not employ generalized states and their motions; instead, the RD is specified using only position $\mu$ and its conjugate momentum $p$ for incoming sensory data $\varphi$.
Environmental objects invoking an agent's sensations can be either static or time dependent, and in turn, the time dependence can be either stationary (not moving on average) or nonstationary.
According to the framework of active inference, the inference of static properties corresponds to passive perception without motor control $a$.
Meanwhile, the inference of time-varying properties renders an agent's active perception of proprioceptive sensations by discharging motor signals $\Delta_\varphi$ via classic reflex arcs.

\begin{figure}[t]
\begin{center}
\includegraphics[width=3.0in]{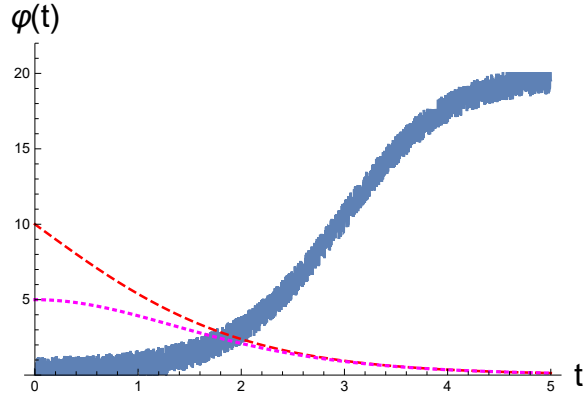}
\caption{Influx of stochastic sensory data $\varphi(t)$ in the blue curve was generated by the environmental process shown in  Eq.~(\ref{temp-record}), which instantly enters the sensory receptor located at the field point $x(t)$.
The dashed curve represents the agent's position at $x(t)$ as a function of time, with its movement starting from $x(0)=10$. The dotted curve represents the magnitude of the latent motor variable $a(t)$ that controls the agent's location. [All curves are in arbitrary units.]}
\label{Fig1}
\end{center}
\end{figure}
In the present simulation, the external hidden state $\vartheta$ is a point property, e.g., temperature or a salient visual feature, which varies with the field point $x$.
As the simplest environmental map, we consider $h(\vartheta,a)=\vartheta(x(a))$ and assume that the sensory influx at the corresponding receptor is given by
\begin{equation}
\label{temp-record}
\varphi = \vartheta + z_{gp},
\end{equation}
where $z_{gp}$ denotes the random fluctuation.
The external property, e.g., temperature, is assumed to display a spatial profile as
\[\vartheta(x)=\vartheta_0/(x^2+1),\]
where $\vartheta_0$ denotes the value at the field origin, and the desired environmental niche is situated at $x=x_d$, where $\vartheta(x_d)=\vartheta_d$.
The biological agent that senses temperature is allowed to navigate through a one-dimensional environment by exploiting the hidden property.
The agent initiates its motion from $x(0)$, where the temperature does not accord with the desired value.
In this case, the agent must fulfill its allostasis at the cost of biochemical energy by exploiting the environment based on
\begin{equation}
\label{agent-action}
x(t)=x(0)+\int_0^t a(t^\prime)dt^\prime,
\end{equation}
where $a(t)$ denotes a motor variable, e.g., agent's velocity.
The nonstationary sensory data $\varphi(t)$ are afferent at the receptor subject to noise $z_{gp}$;
its time dependence is caused by the agent's own motion, i.e., $\varphi(t) = \vartheta(x(a(t)))$, which is assumed to be latent to the agent's brain in the current model.
With the prescribed sensorimotor control, the rate of sensory data averaged over the noise is related to the control variable as
\[
\dot \varphi=\frac{\partial\varphi(x)}{\partial x}a.
\]

The neural observer is not aware of how sensory inputs at the proprioceptor are affected by the motor reflex control of the agent.
In the case of saccadic motor control (Friston et al. 2012b), an agent may stand at a field point without changing its position; however, sampling the salient visual features of the environment through a fast eye movement $a(t)$ makes the visual input nonstationary, i.e., $\varphi(t) = \vartheta(a(t))$.

In Fig.~\ref{Fig1}, we depict streams of sensory data at the agent's receptor as a function of time.
For this simulation, the latent motor variable in Eq.~(\ref{agent-action}) is considered as
\[ a(t)=a(0)\frac{e^t}{(1+e^t)^2},\]
which renders the agent's position in the environment as $x(t)=2x(0)/(1+e^t)$ with $x(0)=-2a(0)$.
For simplicity, we assume that this is hardwired in the agent's reflex pathway over evolutionary and developmental time scales.
The figure shows that the agent, initially located at $x(0)=10$, senses an undesirable stimulus $\vartheta(0)=0.2$; accordingly, it reacts by using motor control to determine an acceptable ambient niche.
For this illustration, we assumed the environmental property at the origin to be $\vartheta_0=20$.
After a period of $\Delta t=5$, the agent finds itself at the origin $x=0$, where the environmental state is marked by the value $\vartheta=20$.

Having prescribed the nonstationary sensory data, we now set up the BM to be integrated by applying Eqs.~(\ref{HamEq1}) and (\ref{HamEq2}) to the generative models below.
We assume that the agent has already learned an optimal generative model; therefore, the agent retains prior expectations regarding the observations and dynamics.
Here, for the demonstration, we consider the learned generative model in its simplest linear form
\begin{eqnarray}
&& g(\mu)=\mu,\label{gmap} \\
&& f(\mu)=-(\mu-\vartheta_d). \label{gfunction}
\end{eqnarray}
Note that the motor control $a$ is not included in the generative model, and the desired sensory data $\vartheta_d$, e.g., temperature, appear as the brain's prior belief of the hidden state.
Accordingly, Eqs.~(\ref{HamEq1}) and (\ref{HamEq2}) are reduced to a coupled set of differential equations for the brain variable $\mu$ and its conjugate momentum $p$ as
\begin{eqnarray}
\dot\mu &=& -(\mu-\vartheta_d) + \frac{1}{m_w}p + \alpha\Delta_\varphi, \label{RD1}\\
\dot p &=& (1+\beta)p -\left(1-\gamma^2\right)\Delta_\varphi. \label{RD2}
\end{eqnarray}
Parameters $\alpha$, $\beta$, and $\gamma$ are proportional to the correlation $\rho$; see Eq.~(\ref{parameters}).
Hence, they become zero when the neural response to the sensory inputs is uncorrelated with neural dynamics, which is not the case in general.
Time-dependent driving terms appearing on the RHS of both equations, namely Eqs.~(\ref{RD1}) and (\ref{RD2}), include the sensorimotor signal $\Delta_\varphi(\mu;\varphi(t))$ given in Eq.~(\ref{MotorSignal}).
The motor variable $a$, which drives the nonstationary inputs $\varphi(t)$, is unknown to the neural observer in our implementation.

\begin{figure*}[t]
\begin{center}
\includegraphics[width=0.75\textwidth]{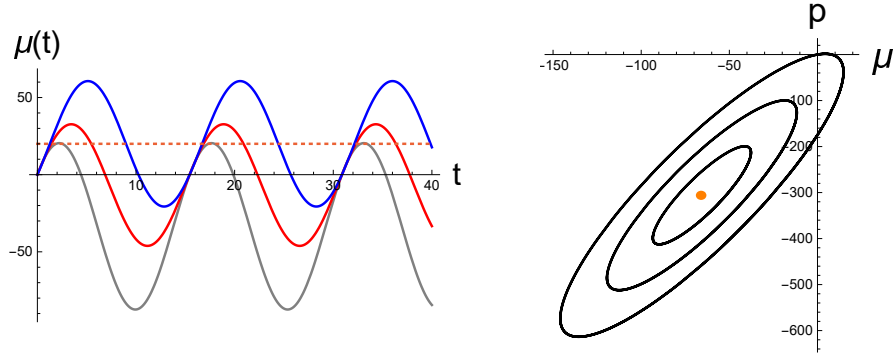}
\caption{Perceptual inference of static sensory data: (Left) Oscillatory brain variables $\mu=\mu(t)$ in time $t$ developed from a common spontaneous state $(\mu(0),p(0))=(0,0)$ by responding to sensory inputs $\varphi= 10$ (gray), $15$ (red), and $20$ (blue). The horizontal dotted line indicates the agent's prior belief regarding sensory input. (Right) Limit cycles in the perceptual state space from an input $\varphi=4.0$ for three initial conditions $(\mu(0),p(0))=(0,0),\ (-20,-100),\ {\rm and}\ (-40,-200)$; where the state space is spanned by continuous brain state $\mu$ and its conjugate momentum $p$ variables. The common fixed point is indicated by an orange bullet at the center of the orbits, which predicts the sensory cause incorrectly.
[All curves are in arbitrary units.]}
\label{Fig2}
\end{center}
\end{figure*}
In the following, for a compact mathematical description, we denote the brain's perceptual state as a column vector
\[
\Psi \equiv
\left(\begin{array}{c}
\mu  \\
p
\end{array}\right).
\]
Vector $\Psi$ represents the brain's current expectation $\mu$ and the associated prediction error $p$ with respect to the sensory causes, as encoded by the neuronal activities performed when encountering a sensory influx.
Therefore, in terms of perceptual vector $\Psi$, Eqs.~(\ref{RD1}) and (\ref{RD1}) are expressed as
\begin{equation}\label{linearRD}
\frac{d\Psi}{dt}+{\cal R}\Psi={\cal S},
\end{equation}
where relaxation matrix ${\cal R}$ is defined as
\begin{equation}\label{relaxation}
{\cal R}
=
\left(\begin{array}{cc}
1+\beta & -\frac{1}{m_w}  \\
(\gamma^2-1)m_z & -(1+\beta)
\end{array}\right),
\end{equation}
and source vector ${\cal S}$ encompassing the sensory influx $\varphi(t)$ is defined as
\begin{equation}\label{source}
{\cal S}
=
\left(\begin{array}{c}
\vartheta_d+\beta\varphi  \\
(\gamma^2-1)m_z\varphi
\end{array}\right).
\end{equation}
Unless it is a pathological case, the steady-state (or equilibrium) solution $\psi_{eq}$ of Eq.~(\ref{linearRD}) is uniquely obtained as
\begin{equation}
\label{fixed-pt}
\Psi_{eq} = {\cal R}^{-1}{\cal S} \equiv
\left(\begin{array}{c}
\mu_{eq}  \\
p_{eq}
\end{array}\right).
\end{equation}
We find it informative to consider the general solution $\Psi(t)$ of Eq.~(\ref{linearRD}) with respect to the fixed point $\psi_{eq}$ by setting
\[
\psi(t)\equiv \Psi(t)-\Psi_{eq}.
\]
To this end, we seek time-dependent solutions for the shifted measure $\psi(t)$ as follows
\[
\frac{d\psi}{dt}+{\cal R}\psi=\delta{\cal S},
\]
where $\delta{\cal S}={\cal S}(t)-{\cal S}(\infty)$.
It is straightforward to integrate the above inhomogeneous differential equation to obtain a formal solution, which is given by
\begin{equation}
\label{formal-sol1}
\psi(t) = e^{-{\cal R}t}\psi(0) + \int_0^t dt^\prime e^{-{\cal R}(t-t^\prime)}\delta{\cal S}(t^\prime).
\end{equation}
Note that $\delta{\cal S}$ becomes zero identically for static sensory inputs; therefore, the relaxation admits simple homogeneous dynamics.
In contrast, for time-varying sensory inputs, the inhomogeneous dynamics driven by the source term is expected to be predominant.
However, on time scales longer than the sensory-influx saturation time $\tau$, it can be shown that $\delta{\cal S}\rightarrow 0$; for instance, $\tau=5$ in Fig.~\ref{Fig1}.
Therefore, for such a time scale, the inhomogeneous contribution in the relaxation diminishes even for time-varying sensory inputs, and the homogeneous contribution is dominant for further time-development.
The ensuing homogeneous relaxation can be expressed in terms of eigenvalues $\lambda_l$ and eigenvectors $\xi^{(l)}$ of the relaxation matrix ${\cal R}$ as
\begin{equation}\label{home-sol}
\psi(t)=\sum_{l=1}^2 c_l e^{-\lambda_l t}\xi^{(l)},
\end{equation}
where expansion coefficients $c_l$ are fixed by initial conditions $\psi(0)$.
The initial conditions $\psi(0)$ represent a spontaneous or resting cognitive state.
In Eq.~(\ref{home-sol}), eigenvalues and eigenvectors are determined by the secular equation
\begin{equation}
\label{secular}
{\cal R}\xi^{(l)}=\lambda_l\xi^{(l)}.
\end{equation}
Then, the solution for the linear RD Eq.~(\ref{linearRD}) is given by
\begin{equation}
\label{formal-sol2}
\Psi(t) = \Psi_{eq} + \sum_{l=1}^2 c_l e^{-\lambda_l t}\xi^{(l)},
\end{equation}
which is exact for perceptual inference, and legitimate for active inference on timescales $t> \tau$.

Before presenting the numerical outcome, we first inspect the nature of fixed points by analyzing the eigenvalues of the relaxation matrix ${\cal R}$ given in Eq.~(\ref{relaxation}).
First, it can be seen that the trace of ${\cal R}$ is zero, which indicates that the two eigenvalues have opposite signs, i.e., $\lambda_1=-\lambda_2$.
Second, the determinant of ${\cal R}$ can be calculated as
\[
{\rm Det}({\cal R})=-\left(1+\beta\right)^2 +\left(\gamma^2-1\right)\frac{m_z}{m_w}.
\]
Therefore, if the correlation $\phi\rightarrow 0$, it can be conjectured that both eigenvalues are real.
This is because ${\rm Det}({\cal R})=\lambda_1\lambda_2\rightarrow -1-m_z/m_w <0$, which yields $\lambda_1^2=\lambda_2^2>0$ using the first conjecture.
Thus, we can conclude that the two eigenvalues are real and have opposite signs.
Therefore, for $\phi=0$, the solution is unstable.
In contrast, when the correlation is retained, ${\rm Det}({\cal R})$ can be positive for a suitable choice of statistical parameters, namely $m_w$, $m_z$, and $\phi$.
In the latter case, the condition $\lambda_1\lambda_2>0$ renders $\lambda_l^2<0$ for both $l=1,2$.
Accordingly, $\lambda_1$ and $\lambda_2$ that have opposite signs are purely imaginary, which makes the fixed point $\Psi_{eq}$ a \textit{center} (Strogatz 2015).
If we define $\lambda_{1,2}\equiv \pm i \omega$, the long-time solution of RD with respect to $\Psi_{eq}$ is expressed as
\[
\psi(t) = c_1 e^{i\omega t}\xi^{(1)} + c_2 e^{-i\omega t}\xi^{(2)},
\]
which specifies a limit cycle with angular frequency $\omega$.
Thus, according to our formulation, the effect of correlation on the brain's RD is not a subsidiary but a crucial component.
Below, we consider numerical illustrations with finite correlation.

We exploited a wide range of parameters for numerically solving Eqs.~(\ref{RD1}) and (\ref{RD2}) and found through numerical observation that there exists a narrow window in the statistical parameters $\sigma_w$, $\sigma_z$, and $\phi$, within which a stable trajectory is allowed for a successful inference.
This finding implies that the agent's brain must learn and hardwire this narrow parameter range over evolutionary and developmental timescales; namely, generative models are conditioned on an individual biological agent.
We denote the instantaneous cognitive state as $(\mu(t),p(t))$ for notational convenience.

In Fig.~\ref{Fig2}, we depict the numerical outcome from the perceptual inference of static sensory inputs.
To obtain the results, we select a particular set of statistical parameters as
\[ \sigma_w=1.0,\ \sigma_z=10,\ {\rm and}\ \phi=-2.8, \]
which specify the neural inertial masses
\[ m_w=4.6\ {\rm and}\ m_z=0.1\times m_w\]
and the coefficients that enter the RD, namely
\[\alpha= -0.60, \beta=-0.28, {\rm and}\ \gamma=-2.8\ (\kappa=10).\]
In Fig.~\ref{Fig2}Left, we depict the brain variable $\mu$ as a function of time, which represents the cognitive expectation of a registered sensory input under the generative model [Eq.~(\ref{gmap})] for three values, namely $\varphi= 10,\ 15, {\rm and}\ 20$.
For all illustrations, the agent's prior belief with regard to the sensory input is set as
\[\vartheta_d=20,\]
which is indicated by the horizontal dotted line.
The blue curve represents the case in which sensory data are in line with the belief.
The RD of the perceptual inference delivers an exact output $(\mu_{eq},p_{eq})=(20,0)$; where $\mu_{eq}$ and $p_{eq}$ are the perceptual outcome of the sensory cause and its prediction error, respectively.
Note that $\mu_{eq}$ and $p_{eq}$ correspond to the temporal averages of $\mu(t)$ and $p(t)$, respectively.
The other two inferences underscore the correct answer.
Figure~\ref{Fig2}Right corresponds to the case of a single sensory data $\varphi=4.0$, which the standing agent senses at the field point $x=2$.
The ensuing trajectories from all three initial spontaneous states have their \textit{limit cycles} in the state space defined by $\mu$ and $p$.
We numerically determined the fixed point to be $(\mu_{eq},p_{eq}) = (-65.6,-306)$ and the two eigenvalues of the relaxation matrix ${\cal R}$ to be $(\lambda_1,\lambda_2)= (1.84i,-1.84i)$, which are purely imaginary and have opposite signs.
Again, the perceptual outcome does not accord with the sensory input; it deviates significantly.

\begin{figure}
\begin{center}
\includegraphics[width=3.0in]{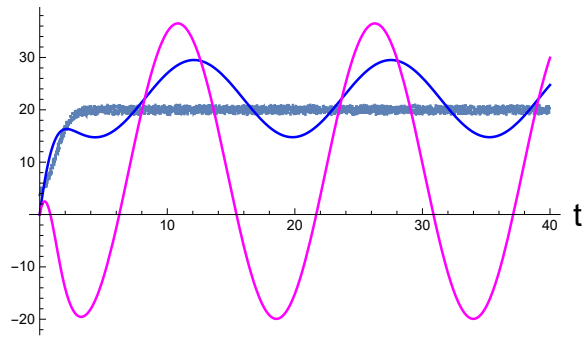}
\caption{Active perception: Time-development of the perceptual state inferring the external causes of sensory inputs altered by the agent's motor control. Blue and magenta curves depict the brain activity $\mu(t)$ and corresponding momentum $p(t)$, respectively. In addition, the noisy curve indicates the nonstationary sensory inputs $\varphi(t)$ entering the sensory receptor at instant $t$. For numerical illustration, we used $\sigma_w=1.0,\ \sigma_z=10,\ {\rm and}\ \phi=-2.8$.  [All curves are in arbitrary units.]}
\label{Fig3}
\end{center}
\end{figure}
\begin{figure}[t]
\begin{center}
\includegraphics[width=3.0in]{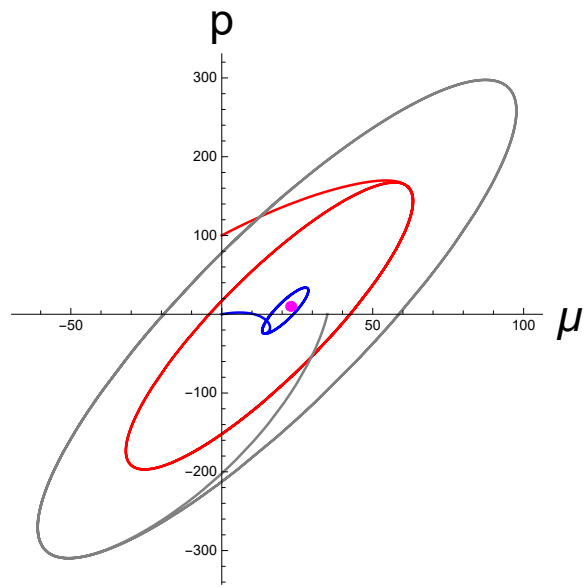}
\caption{Active inference: Temporal development of trajectories rendering stationary limit cycles in the perceptual phase space, spanned by continuous neural state $\mu$ and its conjugate momentum $p$ variables. Data were obtained from the same statistical parameters used in Fig.~\ref{Fig2}. The blue, red, and gray curves correspond to the three initial conditions, $(\mu(0),p(0))=(0,0)$, $(0,100)$, and $(35,0)$, respectively. The angular frequency of the limit cycles is the magnitude of the imaginary eigenvalues of the relaxation matrix ${\cal R}$ given in Eq.~(\ref{relaxation}). The common fixed point is indicated by a magenta bullet at the center of the orbits. [All curves are in arbitrary units.]}
\label{Fig4}
\end{center}
\end{figure}
\begin{figure*}[t]
\begin{center}
\includegraphics[width=0.75\textwidth]{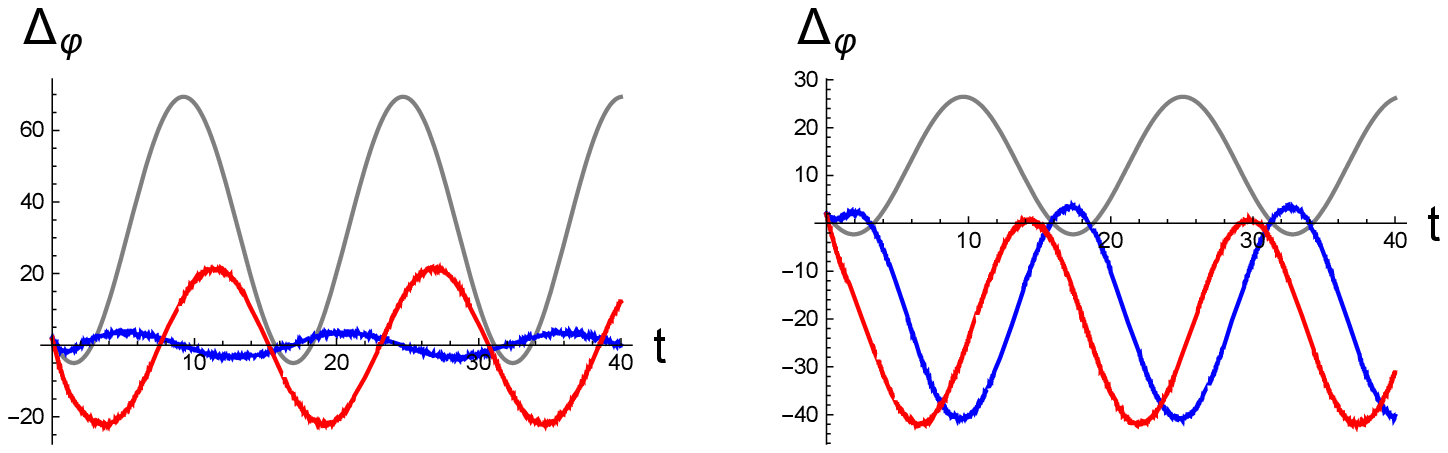}
\caption{Motor signals $\Delta_\varphi(\mu;\varphi(t))$ as a function of time $t$ evoked by the discrepancy between the nonstationary sensory stream and its top-down prediction [Eq.~(\ref{MotorSignal})]. Here, we set the prior belief $\vartheta_d=20$ (Left) and $\vartheta_d=10$ (Right). The blue and red curves represent the results from the initial condition $(\mu(0),p(0))=(0,0)$ and $(0,100)$, respectively. The gray curves represent the corresponding signals from the plain perception of the static sensory input. All data were obtained by setting the statistical parameters as $\sigma_w=1.0,\ \sigma_z=10,\ {\rm and}\ \phi=-2.8$. [All curves are in arbitrary units.]}
\label{Fig5}
\end{center}
\end{figure*}

Next, in Fig.~\ref{Fig3}, we depict the results for active inference, which were calculated using the same generative parameters used in Fig.~\ref{Fig2}.
The agent is initially situated at $x(0)=2$, where it senses the sensory influx $\vartheta(0)=4$, which does not match the desired value $\vartheta_d=20$.
Therefore, the agent reacts to identify a comfortable environmental niche matching its prior belief, which generates nonstationary sensory inputs at the receptors (Fig.~\ref{Fig1}).
The brain variable $\mu$ initially undergoes a transient period at $t\le 5$.
The RD commences from the resting condition $(\mu(0),p(0))=(0,0)$ and then develops a stationary evolution.
Furthermore, we numerically confirmed that the brain's stationary prediction $\mu_{eq}$, which is the brain's perceptual outcome of the sensory cause, is close to but not in line with the prior belief $\vartheta_d$.
The stationary value $p_{eq}$ is estimated to be approximately $8.0$, which is the average of the stationary oscillation of prediction error $p(t)$.

In Fig.~\ref{Fig4}, the trajectory corresponding to that in Fig.~\ref{Fig3} is illustrated in blue in the perceptual state space spanned by $\mu$ and $p$, including two other time developments from different choices of initial conditions.
All data were calculated using the same generative parameters and sensory inputs used for Fig.~\ref{Fig3}.
Regardless of the initial conditions, after each transient period, the trajectories approach stationary limit cycles about a common fixed point, as seen in the case of static sensory inputs in Fig.~\ref{Fig2}Right.
The fixed point $\Psi_{eq}$ and stationary frequency $\omega$ of the limit cycles are not affected by initial conditions, which are solely determined by the generative parameters $m_w$, $m_z$, and $\phi$ and the prior belief $\vartheta_d$ for a given sensory input $\varphi$ [Eqs.~(\ref{fixed-pt}) and (\ref{secular})].
In addition, we numerically observed that the precise location of the fixed points is stochastic, thereby reflecting the noise from the nonstationary sensory influx $\varphi$.

In the framework of active inference, motor behavior is attributed to the inference of the causes of proprioceptive sensations (Adams et al. 2013), and in turn, the prediction errors convey the motor signals in the closed-loop dynamics of perception and motor control.
In Fig.~\ref{Fig5}, we depict the sensorimotor signals $\Delta_\varphi(\mu;\varphi(t))$ that appear as time-dependent driving terms in Eqs.~(\ref{RD1}) and (\ref{RD2}).
In both figures, the agent is assumed to be initially situated such that it can sense the sensory data $\varphi(0)=4$.
After an initial transient period elapses, the motor signals exhibit a stationary oscillation about average zero in Fig.~\ref{Fig5} (Left), implying the successful fulfillment of the active inference of nonstationary sensory influx matching the desired belief $\vartheta_d=20$.
The amplitude of the motor signal shown by the blue curve is smaller than that shown by the red curve, which is also reflected in the size of the corresponding limit cycles in Fig.~\ref{Fig4}.
The prediction-error signal from the plain perception exhibits an oscillatory feature in the gray curve, which arises from the stationary time dependence of the brain variable $\mu(t)$.
The amplitude shows a large variation caused by the significant discrepancy between the static sensory input $\varphi=4$ and its prior belief $\vartheta_d=20$.
In Fig.~\ref{Fig5} (Right), we repeated the calculation with another value: $\vartheta_d=10$.
In this case, the prior belief $\vartheta_d$ regarding the sensory input does not accord with stationary sensory streams.
Therefore, the blue and red signals for active inference oscillate about the negatively shifted values from average zero.
In contrast to Fig.~\ref{Fig5} (Left), the error-signal amplitude of the static input is reduced because the difference between the sensory data and prior belief decreases.

\begin{figure}[t]
\begin{center}
\includegraphics[width=3.0in]{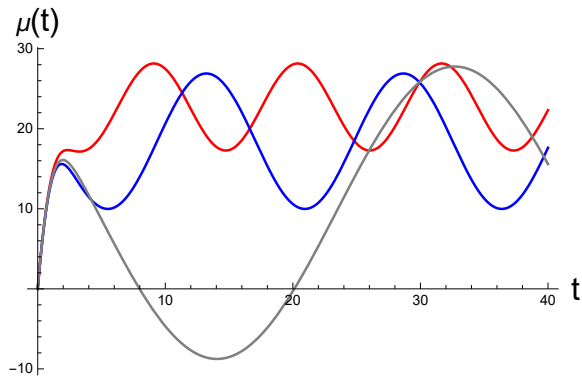}
\caption{Time evolution of the brain variable $\mu$. Here, we vary the correlation $\phi$ for fixed variances $\sigma_w=1.0$ and $\sigma_z=10$. The red, blue, and gray curves correspond to $\phi=-3.0$, $-2.8$, and $-2.6$, respectively. For all data, the agent is environmentally situated at $x=2$, where it senses the transient sensory inputs $\varphi(t)$ induced by the motor reflexes at the proprioceptive level. The agent's initial cognitive state is assumed to be $(\mu(0),p(0))=(0,0)$, and the prior belief is set as $\vartheta_d=20$. [All curves are in arbitrary units.]}
\label{Fig6}
\end{center}
\end{figure}

Next, we consider the role of correlation $\phi$ in the brain's RD, whose value is limited by the constraint $|\phi|\le \sqrt{\sigma_w\sigma_z}$.
To this end, we select three values of $\phi$ for the fixed variances $\sigma_w$ and $\sigma_z$, and we integrate the RD for active inference.
In Fig.~\ref{Fig6}, we present the resulting time evolution of the brain states $\mu$ for the initial condition $(\mu(0),p(0))=(0,0)$.
In this figure, the conjugate momentum variables are not shown.
The noticeable features in the results include the changes in the fixed point and the amplitude of the stationary oscillation with correlation.
The average value of $\mu(t)$ in the periodic oscillation corresponds to the perceptual outcome $\mu_{eq}$ of the sensory data in the stationary limit.
We remark that for all numerical data presented in this work, we selected only negative values for $\phi$.
This choice was made because our numerical inspection revealed that positive correlation does not yield stable solutions.

\begin{figure}[t]
\begin{center}
\includegraphics[width=3.0in]{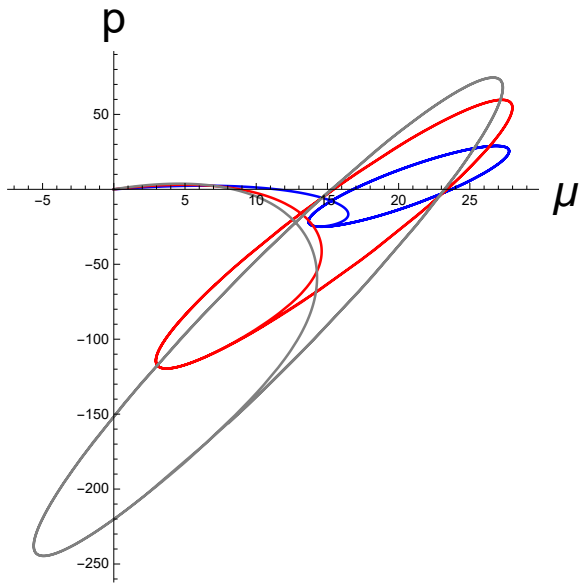}
\caption{Limit cycles in the perceptual phase space spanned by the brain state $\mu$ and its conjugate momentum $p$. Here, we considered several sets of $\sigma_z$ and $\phi$ for a fixed $\sigma_w=1.0$. The red, blue, and gray curves were obtained from $(\sigma_z,\phi)=(50,-6.6)$, $(10,-2.8)$, and $(100,-9.5)$, respectively. For all data, the agent's initial cognitive state is assumed to be $(\mu(0),p(0))=(0,0)$, and the prior belief is set as $\vartheta_d=20$.
The agent is environmentally situated at $x=2$, where it senses the transient sensory inputs $\varphi(t)$ induced by the motor reflexes at the proprioceptive level. [All curves are in arbitrary units.]}
\label{Fig7}
\end{center}
\end{figure}

In Fig.~\ref{Fig7}, as the final numerical manifestation, we show the temporal buildup of the limit cycles in the perceptual phase space; however, this time, we fix $\sigma_w$ while varying $\sigma_z$ and $\phi$.
To generate the red, blue, and gray curves, the tuning parameter $\kappa$ was selected as $\kappa=50,\ 10,\ {\rm and}\ 100$, respectively.
The resulting fixed points are located approximately at the center of each limit cycle, which are not shown.
Similar to that in Fig.~\ref{Fig6}, it can be observed that the positions of the fixed point and amplitudes of oscillation are altered by variations in the statistical parameters.
Evidently, a different set of parameters, namely $\sigma_w$, $\sigma_z$, and $\phi$, which are the learning parameters encoded by the brain, result in a distinctive BM of active inference.

Here, we summarize the major findings from the application of our formulation to a simple nonstationary model.
The brain's BM, i.e., Eqs.~(\ref{RD1}) and (\ref{RD2}), employ linear generative models given in Eqs.~(\ref{gmap}) and (\ref{gfunction}).
\begin{itemize}
\item[(i)] The steady-state solutions of the RD turn out to be a center about which stationary limit cycles (periodic oscillations) are formed as an attractor (Friston and Ao 2012a) in the perceptual phase space, which constitute the brain's nonequilibrium resting states.
\item[(ii)] The nonequilibrium stationarity stems from the pair of purely imaginary eigenvalues of the relaxation matrix with opposite signs, given by Eq.~(\ref{relaxation}); the equal magnitude specifies the angular frequency $\omega$ of the periodic trajectory.
\item[(iii)] Centers are determined by generative parameters and the prior belief for a given sensory input [Eq.~(\ref{fixed-pt})], which represents the outcome of active inference and the entailed prediction error.
\item[(iv)] The theoretical assumption of the statistical dependence of two generative noises describing the brain's expectation of the external dynamics and sensory generation is consequential to ensuring a stationary solution. Furthermore,  based on numerical experience, a negative covariance is necessary for obtaining stable solutions using the current model.
\end{itemize}


\section{Concluding remarks}
\label{conclusion}
In the present study, we continued our effort to make the FEP a more physically principled formalism based on our previous publication (Kim 2018).
We implemented the FEP in the scope of the principle of least action by casting the minimization scheme to the BM described by the effective Hamiltonian equations in the neural phase space.
We deconstructed some of the theoretical details in the first part, which are embedded in the formulation of the FEP, while comparing our approach with other currently prevailing approaches.
In the second half, we demonstrated our proposed continuous-state RD in the Bayesian brain using a simple model, which is biologically relevant to sensorimotor regulation such as motor reflex arcs or saccadic eye movement.
In our theory, the time-integral of the induced IFE in the brain, not the instant variational IFE, performs as an objective function.
In other words, our minimization scheme searches for the tight bound on the sensory uncertainty (average surprisal) and not the instant sensory surprisal.

To present the novel aspects of our formulation, this study focused on the perceptual inference of nonstationary sensory influx at the interface.
The nonstationary sensory inputs were assumed to be unknown or contingent to the neural observer without explicitly engaging in motor-inference dynamics in the BM.
Instead, we considered that the motor signals are triggered by the discrepancies between the sensory inputs at the proprioceptive level and their top-down predictions.
They appeared as nonautonomous source terms in the derived BM, thus completing the sensorimotor dynamics via reflex arcs or oculomotor dynamics of sampling visual stimuli.
This closed-loop dynamics contrasts with the gradient-descent implementation, which involves the double optimization of the top-down belief propagation and the motor inference in message-passing algorithms.
In our present formulation, the sensorimotor inference was not included; however, a mechanism of motor inference can be included explicitly by considering a Langevin equation for a sensorimotor state.
This procedure extends the probabilistic generative model by accommodating the prior density for motor planning for active perception, which is similar to what was done in (Bogacz 2020).

By integrating the Bayesian equations of motion for the considered parsimonious model, we manifested transient limit cycles in the neural phase space, which numerically illustrate the brain's perceptual trajectories performing active perception of the causes of nonstationary sensory stimuli.
Moreover, we revealed that ensuing trajectories and fixed points are affected by the input values of the learning parameters (both diagonal and off-diagonal elements of the covariance matrix) and prior belief regarding sensory data.
The idea of exploring the effect of noise covariance was purely from the theoretical insight without a supporting empirical evidence, which allowed us to drive a stable solution in perceptual and motor-control dynamics.
We did not attempt to explicate in detail the effect of neural inertial masses (precisions) and correlation (noise covariance) on the numerically observed limit cycles.
This was because of the numerical limitation set by the presented model, which permits stable solutions in a significantly narrow window of statistical parameters.
In neurosciences, it is commonly recognized that neural system dynamics implement cognitive processes influencing psychiatric states (Durstewitz et al. 2020).
We hope that the key features of our manifestation will serve to motivate and guide further investigations on more realistic generative models with neurobiological and psychological implications.

Finally, we mention the recent research efforts on synthesizing perception, motor control, and decision making within the FEP (Friston et al. 2015; Friston et al. 2017; Biel et al. 2018; Parr and Friston 2019; van de Laar and de Vries 2019; Tschantz et al. 2020; Da Costa et al. 2020a).
The underlying idea of these studies is rooted in machine learning (Sutton and Barto 1998) and the intuition from nonequilibrium thermodynamics (Parr et al. 2020; Friston 2019), and they attempt to widen the scope of active inference by incorporating prior beliefs regarding behavioral policies.
The new trend supplements the instant IFE to the future expected IFE in a time series, and it formulates the adaptive decision-making processes in action-oriented models.
The assimilation of this feature needs to be studied in depth (Millidge et al. 2020b; Tschantz et al. 2020).
We are currently considering a formulation of motor inference together with the assimilation of extended IFEs in the scope of the least action principle.

\section*{Acknowledgments}
This is a post-peer-review, pre-copyedit version of an article published in Biological Cybernetics. The final authenticated version is available online at: https://doi.org/10.1007/s00422-021-00859-9.
\section*{References}

\end{document}